\documentclass[%
 reprint,
 amsmath,amssymb,
 aps,
]{revtex4-2}

\usepackage{braket}
\usepackage{soul}
\usepackage{graphicx}
\usepackage{dcolumn}

\begin{document}

\title{Spontaneous parametric down-conversion pumped by spatiotemporal structured light}

\author{L. Montenegro}
 \affiliation{Instituto de Física, Universidade de São Paulo, 05315-970 São Paulo, SP, Brazil}
\author{R. F. Barros}%
\email{rafael.fprb@usp.br}
\affiliation{Instituto de Física, Universidade de São Paulo, 05315-970 São Paulo, SP, Brazil}%

\begin{abstract}
Here we investigate the all-optical control of spectral correlations in spontaneous parametric down-conversion. We show that when photon pairs are projected onto high-order spatial modes, the spatial structure of the pump field defines the phase-matching function of the nonlinear interaction.
Thus, by structuring the pump field in both space and spectrum, the biphoton spectral correlations are fully controlled.Considering a standard periodically-poled crystal as the nonlinear medium, we show that the Gouy phase matching method proposed here can generate both spectrally uncorrelated and high-dimensional spectrally entangled photon pairs, similarly to what is achieved with aperiodically-poled crystals. Furthermore, we show that our method can generate a wider class of quantum states if the pump field is a spatiotemporal wavepacket, that is, if its spatial and spectral structures are correlated. 
\end{abstract}

\maketitle

\section{Introduction}
Entangled photon pairs play a fundamental role in modern quantum technologies, notably in quantum communications~\cite{ekert,bennett,tanzilli,yu} and quantum metrology~\cite{Kim,smerzi,kannan}.
The most reliable and efficient way to generate entangled biphoton states is through spontaneous parametric down-conversion (SPDC)~\cite{kwiat,mair,fabian,joshi}, a nonlinear optical process in which photons from a strong pump beam are annihilated, leading to the creation of photon pairs with strong correlations in space, time, and polarization.
These correlations stem from the constraints of the parametric nonlinear interaction, such as the conservation of energy, linear momentum, and orbital angular momentum, and they are defined by both the material properties of the nonlinear medium and the structure of the pump field in all its degrees of freedom~\cite{couteau2018spontaneous}.  

Generally, the optimal use of photon pairs requires accurate control of their quantum correlations. For example, polarization correlations are manipulated by using nonlinear interference, be it with multiple nonlinear crystals~\cite{kwiat1999ultrabright} or by placing a single crystal inside an interferometer~\cite{wong}. Spatial correlations are manipulated by the transfer of angular spectrum from the pump field, which is ensured by the conservation rules of linear and angular momentum in the nonlinear process~\cite{zela,walborn}. The conservation of orbital angular momentum (OAM)~\cite{mair2001entanglement,kopf2025conservation} is particularly important in this context, as photons with entangled OAM can be used for high-dimensional quantum communications (HDQC) in free space~\cite{mirhosseini2015high}.

Recently, the temporal degree of freedom of light has been gaining attention as an alternative high-dimensional basis for quantum communications~\cite{silberhorn,harder,raymer}.
While spatial modes and the OAM of light belong to high-dimensional state spaces, they are not compatible with standard single-mode fibers and are sensitive to turbulence in free-space channels.
On the other hand, broadband spectral/temporal modes~\cite{raymer} are eigenmodes of second-order dispersive media, such as standard single-mode fibers at telecom wavelengths, and thus are promising candidates for high-dimensional quantum communications with the current infrastructures. However, this versatility comes at the cost of more complex measurement schemes, typically involving optical gating in dispersion-engineered nonlinear media~\cite{brecht,ansari,serino}.
Therefore, developing efficient means to generate, control, and measure entangled photon pairs in temporal/spectral modes will be paramount for quantum-secure communications in the future.

In SPDC, spectral correlations between photon pairs depend on two control parameters. First, the pump frequency spectrum controls the frequency-sum correlations due to the conservation of energy.
Second, the frequency dispersion of the nonlinear medium constrains the allowed frequencies through the phase-matching function, which manifests the conservation of linear momentum. While the pump spectrum can be readily controlled with commercially available pulse shapers, changing the phase-matching function typically requires nonlinear crystals with aperiodic sequences of alternating ferroelectric domains, also known as aperiodically-poled crystals (APP)~\cite{pietro,arie,ben}. This approach is efficient in generating spectrally separable photon pairs, which are useful for quantum interference experiments and optical quantum computation, and high-dimensional spectrally entangled states, which are valuable for quantum communications~\cite{arie}. The key limitation of this approach is programmability, as each nonlinear crystal has to be designed for a particular set of quantum states.

In this article, we propose an all-optical method to control the spectral properties of SPDC photons using off-the-shelf nonlinear crystals. Our approach relies on two steps: spatially projecting the SPDC photons and preparing the pump field in a coherent superposition of spatial modes consistent with the selection rules of the nonlinear process. By doing so, we exploit the Gouy phase intrinsic to spatially structured light, which leads to mode-dependent phase-matching functions that can be coherently superposed to engineer the desired spectral correlations. We benchmark our method, which we call Gouy phase matching (GPM), by inverse-designing Hermite-Gaussian phase-matching functions, obtaining excellent fidelity up to the third order with superpositions of up to 10 pump modes. Finally, we show that pump fields with correlated spatial and spectral structures, also known as spatiotemporal wavepackets~\cite{kondakci2019optical}, generate a broader class of quantum states inaccessible to the APP crystal approach. These states include, but are not limited to, spectrally separable photons in high-order Hermite-Gaussian temporal modes.

\begin{figure}
    \centering   \includegraphics[width=0.9\linewidth]{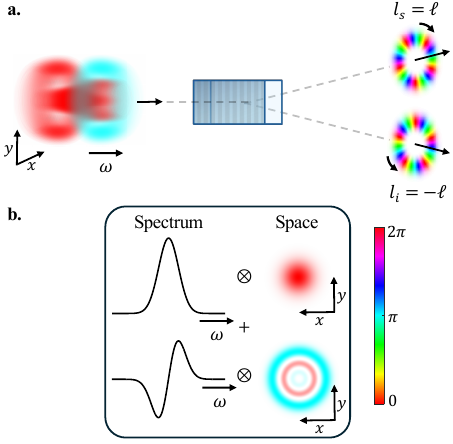}
    \caption{a) Conceptual image of SPDC photons produced by pumping a nonlinear crystal with a spatiotemporally structured pump beam. The pump field shown is composed of ultrafast pulse shapes attached to superpositions of radial Laguerre-Gaussian spatial modes $(l_p = 0)$, as illustrated in b).  The signal and idler photons are spatially projected onto counter-rotating OAM modes with $l_s =\ell$ and $l_i = -\ell$.}
    \label{Fig:Concept}
\end{figure}

\section{Fundamentals}
\subsection{Mode-dependent Joint-Spectral Amplitude (mJSA)}
In SPDC, a nonlinear crystal is illuminated by a bright pump \textit{(p)} field, driving the creation of pairs of photons in signal (\textit{s}) and idler (\textit{i}) modes. 
These modes are initially in a vacuum state and evolve according to the following interaction Hamiltonian
\begin{equation}\label{hamiltonian}
    \mathcal{H}_{int}(t) = \int d^3r\,\, d(z){E}_p(\vec{r},t)\hat{E}_{s}(\vec{r},t)\hat{E}_{i}(\vec{r},t)\,,
\end{equation}
where $d(z)$ is an effective nonlinear susceptibility, potentially non-uniform across the nonlinear medium.
The pump field $E_p(\vec{r},t)$ is taken as a classical coherent wave propagating in the $z$ direction
\begin{equation}\label{field}
        E_{p}(\vec{r},t) =\sum_{l_{p},m_{p}}\int \, d\omega_{p}\;\alpha^{l_p}_{m_p}(\omega_p)\,u^{l_{p}}_{m_{p}}(\vec{r})\,e^{i(k_{p}z-\omega_{p}t)} + c.c.\,,
\end{equation}
where $\alpha^{l_p}_{m_p}(\omega_p)$ is the amplitude of a monochromatic wave with frequency $\omega_p$ and transverse spatial mode $u^{l_{p}}_{m_{p}}(\vec{r})$. Note that this pump field represents an arbitrary nonseparable structure in space and time, i.e., $E_p(\vec{r},t)\neq\mathcal{F}(\vec{r})\mathcal{G}(t)$. On the other hand, the signal and idler field operators $\hat{E}_j(\vec{r},t)$ ($j = s,i$) are given by
\begin{equation}\label{field2}
    \hat{E}_{j}(\vec{r},t) =\sum_{l_{j},m_{j}} \int \, d\omega_{j}\,u^{l_{j}}_{m_{j}}(\vec{r})\,\hat{a}^{l_j}_{m_j}(\omega_{j})e^{i(k_{j}z-\omega_{j}t)} + H.c.\,,
\end{equation}
where $\hat{a}^{l_j}_{m_j}(\omega_j)$ is the annihilation operator of the transverse spatial mode $u^{l_{j}}_{m_{j}}(\vec{r})$ at the frequency $\omega_j$.

In the standard SPDC regime, where the probability of generating a photon pair within a measurement window is small, the SPDC state can be approximated by~\cite{fritz}
\begin{equation}\label{state2}
    \ket{\psi}\! \approx \!\sum_{l_{s},m_{s}}\!\sum_{l_{i},m_{i}}\int d\omega_{s}d\omega_{i}\, \,X^{l_{s},l_{i}}_{m_{s},m_{i}}(\omega_{s},\omega_{i})\hat{a}^{l_s\dagger}_{m_s}(\omega_{s})\hat{a}^{l_i\dagger}_{m_i}(\omega_{i})\ket{0}\,.
\end{equation}
Here, $X^{l_{s},l_{i}}_{m_{s},m_{i}}(\omega_{s},\omega_{i})$ is a {mode-dependent joint spectral amplitude} (mJSA) defined as
\begin{equation}\label{jsa}
\begin{split}
    X^{l_{s},l_{i}}_{m_{s},m_{i}}(\omega_{s},\omega_{i}) &= g\sum_{l_{p},m_{p}} \alpha^{l_{p}}_{m_{p}}(\omega_{s}+\omega_{i})\;\Lambda^{l_{p},l_{s},l_{i}}_{m_{p},m_{s},m_{i}}(0)\times\\ &\times\int^{\frac{L}{2}}_{-\frac{L}{2}} \!dz\,\; d(z)\, \frac{e^{i[\Delta kz -\Delta\phi_G(z)]}}{\sqrt{1+(z/z_{r})^{2}}}\,,
\end{split}  
\end{equation}
\\
where, $g$ is a constant factor, $L$ is the length of the nonlinear crystal, $z_r$ is the Rayleigh length,  and $\Delta k = k_p-k_s - k_i$ is the wave-vector mismatch. The mJSA gives the probability amplitude for a pair of photons in the modes ($l_s,m_s,\omega_s$) and ($l_i,m_i,\omega_i$). The spatial modes involved play a role in two ways. First, the overlap integral
\begin{equation}
    \Lambda^{l_{p},l_{s},l_{i}}_{m_{p},m_{s},m_{i}}(z)= \int d^3r\,\, u^{l_p}_{m_p}(\vec{r})u^{l_s*}_{m_s}(\vec{r})u^{l_i*}_{m_i}(\vec{r})\,,
\end{equation}
defines the effective nonlinear coupling for each combination of spatial modes, as well as the spatial selection rules of the nonlinear process. For example, if we work in the Laguerre-Gaussian (LG) basis, so that $l_j$ and $p_j$  represent the azimuthal and radial indices, the conservation of OAM implies that $l_p = l_s + l_i$. A more general discussion on the spatial selection rules in three-wave mixing processes can be found in~\cite{schwob,bier,walborn}. Second, the phase of each spatial mode evolves upon propagation as 
\begin{equation}
    u^{l_j}_{m_j}(z) = u^{l_j}_{m_j}(0)\exp{\left[ik_jz-i\phi_j(z)\right]}\,,
\end{equation}
where the mode-dependent correction to the wave number is the Gouy phase~\cite{wingful}, defined as $\phi_j(z) = (O_j+1)\tan^{-1}(z/z_r)$ for a mode order $O_j(l_j,m_j) = |l_j| + 2m_j$. Thus, the wave vector mismatch in \eqref{jsa} is accompanied by a Gouy phase mismatch term
\begin{equation}
\begin{split}
    \Delta\phi_G(z) &= \phi_p(z)-\phi_s(z)-\phi_i(z)\\
    &=(O_p-O_s-O_i-1)\tan^{-1}(z/z_R)\,,
    \end{split}
\end{equation}
 resulting in a mode-dependent contribution to the phase-matching function .

\begin{figure}
    \centering    \includegraphics[width=1\linewidth]{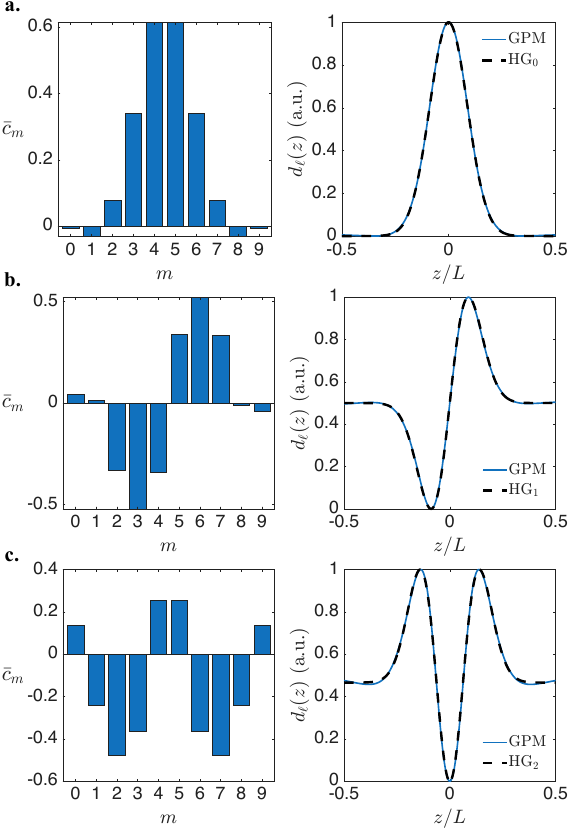}
    \caption{Hermite-Gaussian nonlinear coefficients obtained $L = 15~$mm, $\ell=9$, $\delta\omega \Delta\beta L = 16$, and $z_R/L=0.2$. The first column shows the coefficients $\bar{c}_m$ optimized with a genetic algorithm to minimize the Euclidean distance $||d_\ell(z)-d_T(z)||$ to Hermite-Gaussian targets of \textbf{a.} zeroth, \textbf{b.} first and \textbf{c.} second order. The target nonlinear coefficients and the corresponding optimization results are shown on the second column.}
    \label{optHG}
\end{figure}

\subsection{Proposal for all-optical quasi-phase matching}

{From \eqref{state2}, we see that the spectral correlations of photon pairs projected onto the spatial modes $\{m_s,l_s\}$ and $\{m_i,l_i\}$ are completely defined by the mJSA $X^{l_{s},l_{i}}_{m_{s},m_{i}}(\omega_{s},\omega_{i})$, which depends on the spatial and spectral structures of the pump via the set of amplitudes $\alpha_{m_p}^{l_p}(\omega_s+\omega_i)$. Here we consider the case illustrated in Fig.~\ref{Fig:Concept}, where the pump is prepared in a superposition of radial LG modes with $l_p = 0$ and $m_p =m\geq 0$, and the photons are projected onto OAM modes with $l_s=-l_i=\ell$ and $m_s = m_i = 0$. Besides OAM conservation, the spatial selection rules imply that only the pump modes with $0<m<\ell$ will contribute to the SPDC state~\cite{bier}, i.e., the mJSA is controlled by a set of $\ell+1$ complex coefficients.}

Spectral correlations between signal and idler photons come from two sources. First, conservation of energy imposes $\omega_p = \omega_s+\omega_i$ and therefore, the spectral structure of the pump determines the correlations along the $\omega_+ =\omega_s+\omega_i$ direction. Second, the dispersive properties of the medium and the conservation of linear momentum restrict the allowed frequencies through the mismatch term $\Delta k$.
Considering only linear dispersion, we have $\Delta k = \Delta k_0 + \omega_s(v_{p}^{-1}-v_{s}^{-1}) + \omega_i(v_{p}^{-1}-v_{i}^{-1})$, where $\Delta k_0$ is a frequency-independent term, and $v_{j}$ $(j = p,s,i)$ are the group velocities at the central frequencies.
Here we consider the case of symmetric group velocity matching (sGVM)~\cite{donohue}, which occurs when the pump group velocity lies between the group velocities of signal and idler 
\begin{equation}
    v_{p}^{-1} - v_{s}^{-1} = -(v_{p}^{-1} - v_{i}^{-1}) = \Delta\beta \,.
\end{equation} 
This condition implies that $\Delta k= \Delta k_0 +\Delta \beta (\omega_s-\omega_i)$, meaning that the phase mismatch affects the spectral correlations along the $\omega_-=\omega_s-\omega_i)$ axis, which is orthogonal to the $\omega_+$ axis defined by the pump frequency spectrum. 
Although apparently too restrictive, sGVM occurs naturally at telecom wavelengths in Type-II KTiOPO4 (KTP)~\cite{arie,donohue}, which is the nonlinear medium that we consider in all our simulations.

\begin{figure}
    \centering    \includegraphics[width=0.8\linewidth]{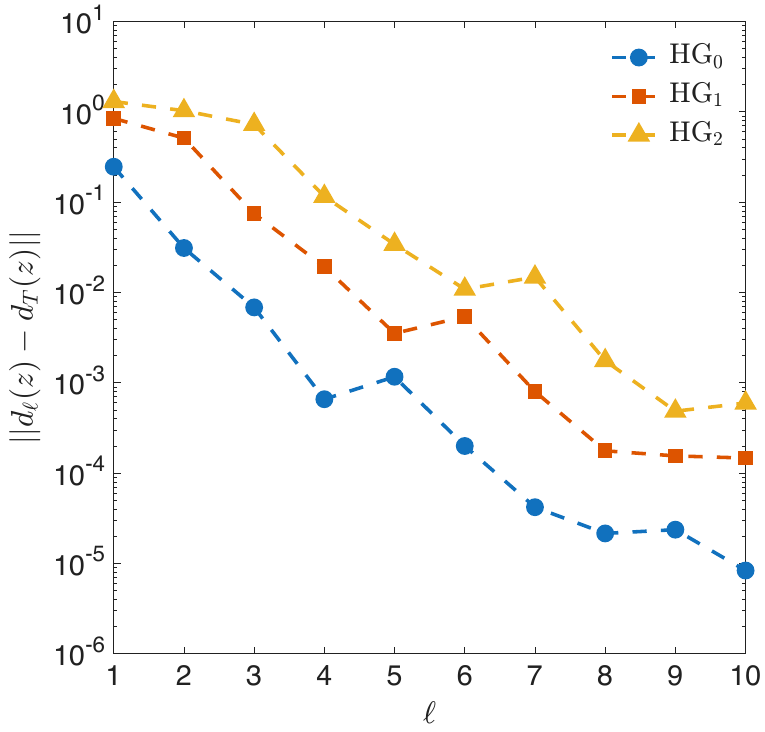}
    \caption{Fidelity of the effective nonlinear susceptibility to Hermite-Gaussian targets obtained by a genetic optimization of the pump spatial structure with varying $\ell$. The fidelity is measured as the Euclidean distance to the target nonlinear coefficient.}
    \label{fidel}
\end{figure}

\begin{figure*}[t]
    \centering
    \includegraphics[width=1\linewidth]{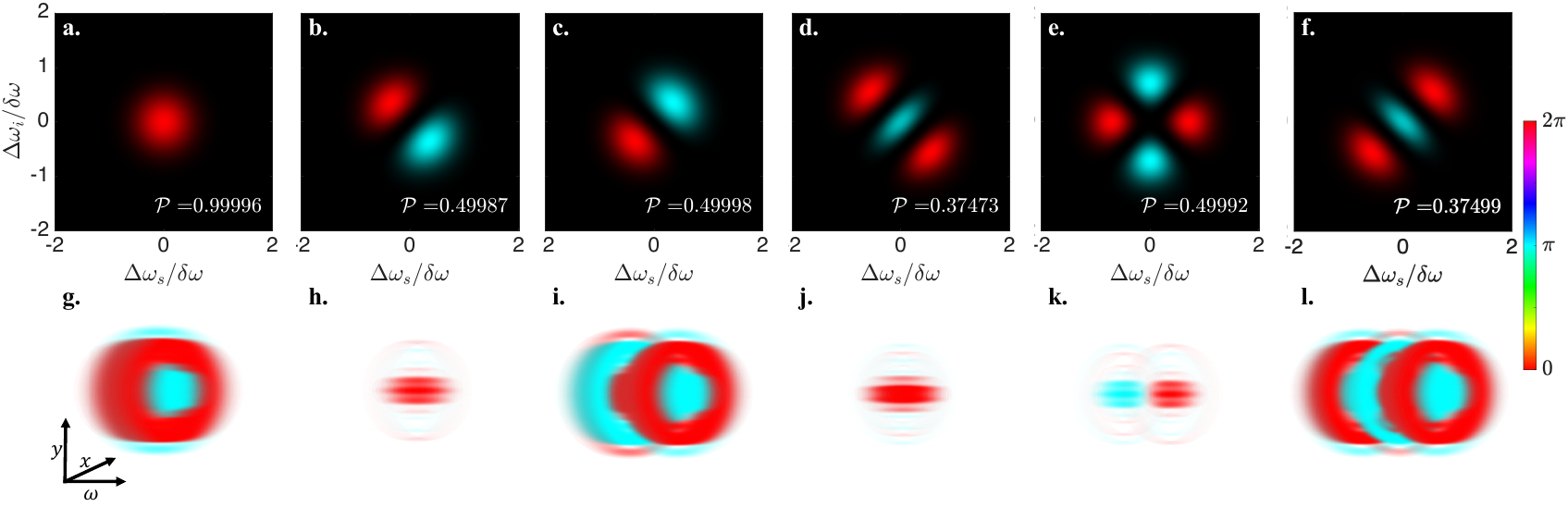}
    \caption{Hermite-Gaussian joint spectral amplitudes in the rotated basis and their respective spatiotemporal pumps. We show the HG$_{00}$ state in \textbf{a.}/\textbf{g.}, HG$_{01}$ in \textbf{b.}/\textbf{h.}, HG$_{10}$ in \textbf{c.}/\textbf{i.}, HG$_{02}$ in \textbf{d.}/\textbf{j.}, HG$_{11}$ in \textbf{e.}/\textbf{k.}, and HG$_{20}$ in \textbf{f.}/\textbf{l.}. Here $\Delta\omega_{s/i} = \omega_{s/i}-\omega^{0}_{s/i}$, $\omega^0$ being the central, phase-matched frequency corresponding to the wavelength of $1560\,$nm. The nonlinear medium considered is a type-II KTiOPO4 crystal, and all other parameters are the same as in Fig.~\ref{optHG}.}
    \label{Fig:SepJSA}
\end{figure*}

With the above conditions satisfied, the mJSA becomes
\begin{equation}\label{JSAnonsep}
\begin{split}
    X^{\ell,-\ell}_{0,0}(\omega_{+},\omega_{-}) &= \sum_{m=0}^\ell\alpha_{m}(\omega_+)\mathcal{F}_{m}^{\ell}(\omega_-)\,,
\end{split}  
\end{equation}
where we defined a mode-dependent phase-matching function (PMF) as
\begin{equation}\label{PMFunction}
    \begin{split}
        \mathcal{F}_{m}^{\ell}(\omega_-) = \int^{\frac{L}{2}}_{-\frac{L}{2}}dz\, d^\ell_{m}(z) e^{i(\Delta \beta \,\omega_-+\Delta k_0)\,z}\,,
    \end{split}
\end{equation}
with
\begin{equation}\label{NLCoeff-eff}
    d^{\ell}_{m}(z) = g \Lambda^{0,\ell,-\ell}_{m,0,0}(0) \,d(z)\,\left[\frac{e^{-i[2(m-\ell)-1]\tan^{-1}(z/z_R)}}{\sqrt{1+(z/z_{r})^{2}}}\right]\,.
\end{equation}
Notice that the term in brackets comes from the Gouy phase mismatch between pump, signal and idler, and that its effect is to create a longitudinal variation of the effective nonlinear coefficient. Importantly, this Gouy phase effect appears on equal footing to the actual $z$-dependent nonlinear susceptibility $d(z)$, whose control requires material processing techniques such as ferroelectric domain engineering~\cite{arie} and nonlinearity erasure~\cite{riazi2019biphoton}. This result suggests that quantum control over the SPDC spectral correlations can also be achieved all-optically through the spatiotemporal manipulation of the pump field. We call this effect Gouy phase matching (GPM).

\section{Results}
\subsection{GPM with a spatially structured pump}

First, we consider the simplest case of a pump field with separable spatial and spectral structures, meaning that all spatial modes have the same frequency spectrum $\alpha_{m_p}(\omega_+) = S(\omega_+)c_{m_p}$. In this case, we obtain simply
\begin{equation}\label{jsa2}
\begin{split}
    X^{\ell,-\ell}_{0,0}(\omega_{+},\omega_{-}) &= S(\omega_+)\mathcal{F}_\ell(\omega_-)\,,
\end{split}  
\end{equation}
where the PMF is given by
\begin{equation}\label{PMseparable}
\begin{split}
    \mathcal{F}_\ell(\omega_-) &=\sum_{m=0}^\ell c_{m}\mathcal{F}_{m}^{\ell}(\omega_-)\,,\\
    &=\int^{\infty}_{-\infty}dz\, d_\ell(z) e^{i\Delta \beta\,\omega_{-}\,z}\,.
    \end{split}
\end{equation}
Notice that we have extended the integration limits to $\pm\infty$, so the effective nonlinear coefficient becomes
\begin{equation}\label{NLCoeff-separable}
    d_{\ell}(z) = \frac{\Pi(z)\bar{d}(z)}{\sqrt{1 + (z/z_R)^2}}\sum_{m = 0}^\ell \Bar{c}_{m} \,{e^{-i\left(2m-\ell\right)\tan^{-1}(z/z_R)}}\,,
\end{equation}
where $\Bar{c}_{m} = g\Lambda^{0,\ell,-\ell}_{m,0,0}(0)c_{m}$, $\Pi(z)$ is a square window function which is 1 for $-L/2\leq z\leq L/2$ and 0 otherwise, and
\begin{equation}
    \bar{d}(z) = d(z)e^{i\left[\Delta k_0 z +(\ell+1)\tan^{-1}(z/z_R)\right]}\,.
\end{equation}
Without loss of generality, we will consider that the nonlinear medium is an off-the-shelf periodically poled crystal that is quasi-phase matched so that $\bar{d}(z)=d_0$. 

\begin{figure*}
    \centering
    \includegraphics[width=1\linewidth]{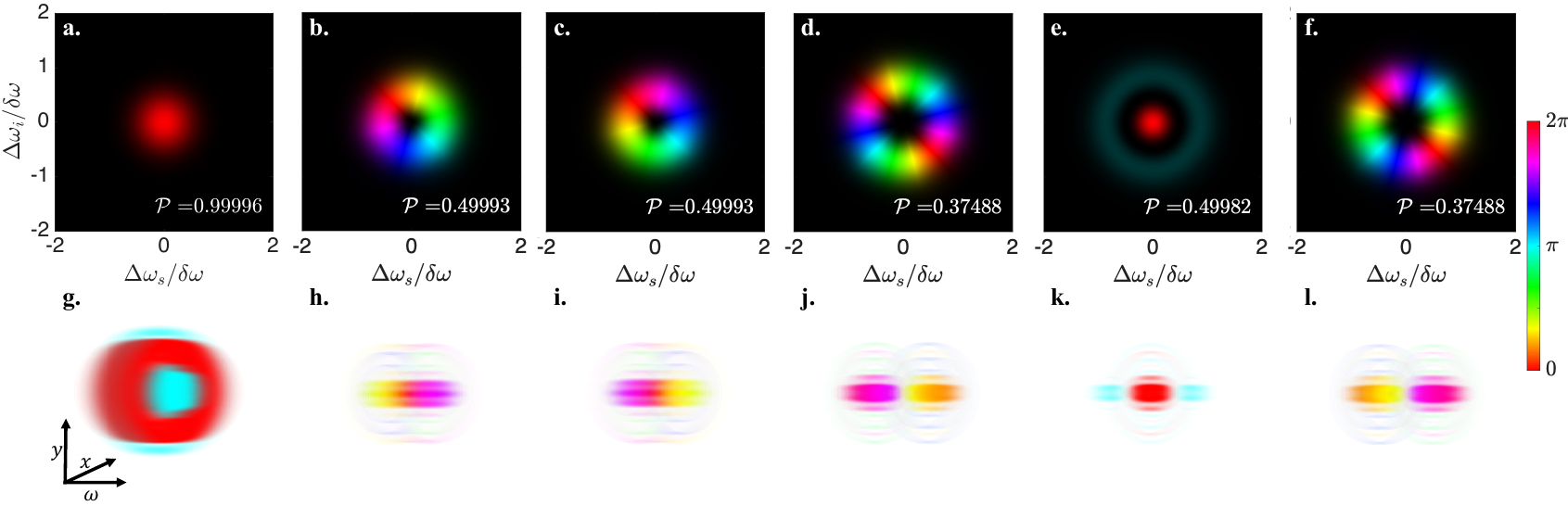}
    \caption{Joint spectral amplitudes and their respective spatiotemporal pump structures in the Laguerre-Gauss basis. These states represent superpositions of the states created in \ref{Fig:SepJSA} and are all maximally entangled. Figures \textbf{a.} and \textbf{g.} represent the gaussian mode; figures \textbf{b.} and \textbf{h.} and \textbf{c.} and \textbf{i.} represent modes $LG^{+1}_{0}$ and $LG^{-1}_{0}$, respectively, while figures \textbf{d.} and \textbf{j.}, \textbf{e.} and \textbf{k.} and \textbf{f.} and \textbf{l.} represent the modes of order two, $LG^{+2}_{0}$, $LG^{0}_{1}$ and $LG^{-2}_{0}$.}
    \label{CorrJSA}
\end{figure*}

As can be seen from Eqs.(\ref{PMseparable}) and (\ref{NLCoeff-separable}), the PMF is a scaled Fourier transform of the effective nonlinear coefficient, as observed in \cite{arie}. Thus, a target PMF $\mathcal{F}_T(\omega_-)$ can be obtained by inverse-designing the spatial structure of the pump field so that
\begin{equation}\label{NLcoeff_Fourier}
    d_\ell(z) = \int_{-\infty}^{\infty} d\Bar{\omega}_- \mathcal{F}_T(\omega_-)e^{-i\Delta\beta\,\omega_-\,z}\,.
\end{equation}
Notable examples of target PMFs are the Hermite-Gaussian (HG) spectral modes, defined as
\begin{equation}
    \mathcal{F}_T(\omega_-) = \textrm{H}_m\!\left( \frac{\sqrt{2}\omega_-}{\delta\omega}\right)e^{-\omega_-^2/\delta\omega^2}\,,
\end{equation}
where $\textrm{H}_m$ is the Hermite polynomial of order $m$, and $\delta\omega$ is the bandwidth defining the mode basis. From \eqref{NLcoeff_Fourier}, we obtain that the required nonlinear coefficient is
\begin{equation}
    d_{T}(z) = \textrm{H}_m\left(\frac{\delta\omega\,\Delta\beta\,z}{\sqrt{2}}\right)e^{-(\delta\omega\,\Delta\beta\, z)^2/4}\,,
\end{equation}
where the bandwidth must be $\delta\omega \lesssim 2/L\Delta\beta$ to ensure good confinement to the length of the nonlinear crystal. $d_\ell(z)$ itself is a combination of oscillatory terms connected to the Gouy phase, yielding a representation analogous to that of a Fourier series truncated at $m=\ell$ and with period $\approx 2\pi z_r$. Hence, by preparing the pump in an adequate superposition, we can build any well behaved function, its accuracy depending on the number of modes included.

In Fig.~\ref{optHG} we show examples of Hermite-Gaussian PMFs obtained with GPM. The results were obtained with fixed parameters $L = 15~$mm, $\ell=9$, $\delta\omega \Delta\beta L = 16$, and $z_R/L=0.2$, with coefficients $\bar{c}_m$ optimized with a genetic algorithm to minimize the Euclidean distance $||d_\ell(z)-d_T(z)||$. The results show that using a superposition of $\ell+1=10$ pump modes, the GPM can be used to generate Hermite-Gaussian PM functions of up to second-order with high fidelity. In Fig.~\ref{fidel} we show how the performance of the GPM varies with $\ell$, and hence with the number of pump modes available for the optimization. We see that higher-order PM functions require more pump modes to be generated accurately, as expected given their faster-oscillating structures. However, an optimal Gaussian PM function, which is desirable for the generation of spectrally pure photon pairs, can be obtained with only 8 pump modes.

In Fig. \ref{Fig:SepJSA}, we show examples of Hermite-Gaussian JSAs obtained by preparing both the pump spectrum and the PMF with Hermite-Gaussian modes up to the second order. As the pump and the PMF dictate the structure of the JSA along the $\omega_+$ and $\omega_-$ directions, the JSAs obtained resemble standard Hermite-Gaussian modes rotated by 45$^\circ$, which are completely separable in the $(\omega_+,\omega_-)$ basis. Consequently, in the $(\omega_s,\omega_i)$ basis, only the Gaussian JSA is separable, with all higher-order JSAs featuring strong spectral correlations. We confirm this result by calculating the spectral purity $\mathcal{P}$, taken as the purity of the unheralded signal and idler photons. As we shown in the figure, HG JSAs of first order have $\mathcal{P}\approx 0.5$, as expected from maximally entangled biphoton states of dimension 2. For second-order JSAs, we see $\mathcal{P} \leq 0.5$, indicating that these quantum states have some degree of high-dimensional entanglement. All these results are on par with those obtained with the SPP approach~\cite{arie}. These results are on par with those obtained with the APP approach~\cite{arie}, with the advantage that all the JSAs shown were simulated using a single periodically-poled crystal.

\subsection{GPM with a spatiotemporally structured pump}

\begin{figure}
    \centering
    \includegraphics[width=0.9\linewidth]{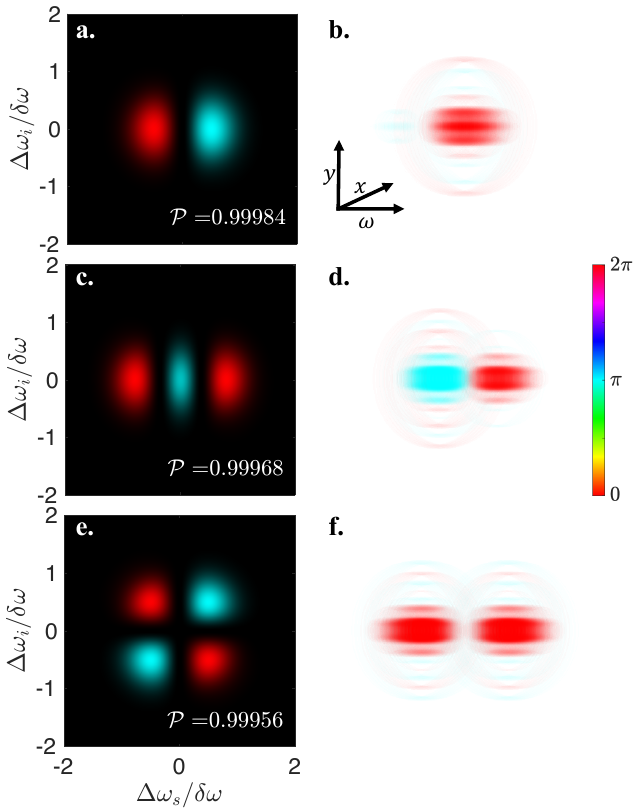}
    \caption{In \textbf{a.}, \textbf{c.} and \textbf{e.}, we show the joint spectral amplitudes of separable photon pairs in high-order Hermite-Gaussian spectral modes. The corresponding spatiospectral structures of the pump field are shown in \textbf{b.}, \textbf{d.}, and \textbf{f.}.}
    \label{septotal}
\end{figure}

So far we have discussed the control of the JSA by preparing the pump spectrum and, separately, tailoring either its spatial structure or the nonlinear crystal. As it is evident from \eqref{jsa2}, these methods have a common limitation: they can only generate JSAs that are separable in the rotated $(\omega_+,\omega_-)$ frame. Such separable JSAs cover all possible Hermite-Gaussian joint spectra of the form HG$_m(\omega_+)$HG$_n(\omega_-)$, forming a basis in terms of which any JSA can be represented.

Given the completeness of the Hermite-Gaussian mode basis, genuine two-dimensional control of the biphoton spectrum can be achieved by superposing $(\omega_+,\omega_-)$-separable JSAs. With the APP approach, an arbitrary JSA requires a set of different aperiodically poled crystals combined in an interferometric scheme, each pumped with a different HG pump spectrum. Evidently, this method scales unfavorably, requiring intense resource overheads for JSAs of increasing complexity. On the other hand, using GPM, the same arbitrary JSA requires a superposition of pump spectra, each attached to a different spatial structure - a spatiotemporal wavepacket. In this case, the joint spectrum is described by the general mJSA of \eqref{JSAnonsep}. 

In Fig. \ref{CorrJSA} we show how the method works by generating biphoton states in superpositions of Hermite-Gaussian mJSAs of the same order. These superpositions stem from the transformation between Hermite-Gaussian and Laguerre-Gaussian modes in the spatial domain, which is the reason for the cylindrical symmetry of their intensity profiles and for the presence of phase singularities. Interestingly, these topological structures exist in the spectral correlations between photons rather than in the spatial structure of the light field. Taking this parallel further will be a topic of future work, but the recent advances in topological photonics~\cite{rubinsztein2017roadmap} suggest that such quantum states might be robust to perturbations caused by propagation in dispersive media.

In fact, any superposition of the mJSAs shown in Fig.~\ref{Fig:SepJSA} can be created by superposing their pump fields with the corresponding amplitudes. An important case is that of high-order spectrally separable states. From the mJSAs, separable states can be identified as HG structures oriented along the $(\omega_s,\omega_i)$ axes, meaning that the signal and idler photons have independent HG spectral modes. We show in Fig.~\ref{septotal} some examples of separable JSAs and the pump fields that generate them, corresponding to signal and idler in HG modes up to the 2nd order. These results highlight a potential application for GPM in quantum communication and information processing, given that the preparation of single photons in high-order Hermite-Gaussian modes normally requires lossy spatial light modulators. Although such states can be generated from separable photons with Gaussian spectra, their conversion to high-order HG modes requires external modulation, hindering the applicability of this approach to loss-sensitive applications.

\section{Conclusion}
In summary, we have investigated the all-optical control of biphoton spectral correlations through what we called Gouy phase matching. We have shown that the spatiospectral manipulation of the pump field results in genuine two-dimensional control of the joint spectral amplitude, allowing the accurate preparation of both separable and entangled quantum states. Importantly, our method is field-programmable in the sense that it does not rely on crystal engineering. Furthermore, our results show that when the pump field is a spatiotemporal wavepacket, Gouy phase matching allows the preparation of quantum states that cannot be achieved by a single aperiodically poled crystal, including, but not limited to, spectrally separable photons in high-order temporal modes. Our results pave the way for the use of spatiotemporal structured light in quantum optics, with clear potential for applications in high-dimensional quantum communications.

\section*{Acknowledgements}
The authors thank Robert Fickler and Qiwen Zhan for valuable discussions. This work was supported by FAPESP (São Paulo Research Foundation) under Grants No. 2025/00093-6 and 2024/08450-0.

\bibliography{sample}

\end{document}